\documentstyle[aps]{revtex} 
\begin{document} 
\draft 
\preprint{hep-th 0110198} 
\title{Black hole entropy: classical and quantum aspects\footnote{Based
on lectures given at the YATI Conference on Black Hole Astrophysics,
Kolkata, India, April 2001.}}
\author{Parthasarathi Majumdar\footnote{email: partha@imsc.ernet.in}} 
\address{The Institute of Mathematical Sciences, CIT Campus, Madras
600113, India.} 
\maketitle
\begin{abstract} 
An elementary introduction is given to the problem of black hole entropy
as formulated by Bekenstein and Hawking, based on the so-called Laws of
Black Hole Mechanics. Wheeler's `It from Bit' picture is presented as
an explanation of plausibility of the Bekenstein-Hawking Area Law. A
variant of this picture that takes better account of the symmetries of
general relativity is shown to yield corrections to the Area Law that are
logarithmic in the horizon area, with a finite, fixed coefficient.
The Holographic hypothesis, tacitly assumed in the above
considerations, is briefly described and the beginnings of a general proof
of the hypothesis is sketched, within an approach to quantum gravitation
which is non-perturbative in nature, namely Non-perturbative Quantum
General Relativity (also known as Quantum Geometry). The holographic
entropy bound is shown to be somewhat tightened due to the corrections
obtained earlier. A brief summary of Quantum Geometry  approach is
included, with a sketch of
a demonstration that precisely the log area corrections obtained from the
variant of the It from Bit picture adopted  earlier emerges for the
entropy of generic black holes within this formalism.  
\end{abstract}

\section{Introduction}

We begin with the startling fact, following basically from Newton's law of
gravitation, that if all matter (and radiation) on earth were to be
squeezed into an ordinary marble of a couple of centimetres in diameter,
the gravitational field of this marble would be so strong as to render it
invisible (or `black').  While this is an interesting speculation,
it actually portrays realistic situations. Consider the death
of stars after the nuclear fuel that powers them is exhausted; the star
dies, often with a spectacular explosion, and the corpse begins to
contract due to its own gravitational attraction, in the absence of
outward pressures generated by nuclear explosions that existed during its
lifetime. If such a star-remnant weighs more than a couple of times the
mass of the sun, the internal gravitational attraction supercedes all
other repulsive forces (like Pauli degeneracy pressure) to cause it to
collapse indefinitely under its own weight. During this gravitational
collapse, the star contracts, and its density rises. A stage is reached
when the resultant gravitational force is so strong as to trap everything,
including light to the stellar surface. The particular size of the star
for which this happens is given in terms of its remnant mass and is known
as the Schwarzschild radius $r_S = 2GM/c^2$. It is as though the surface
of the star at this stage is a one-way membrane - matter and energy can go
in through it but nothing can ever come out.

The collapse of the star under its own weight continues beyond
this stage, until all the matter (and radiation) has collapsed {\it to a
single point} ! Of course this is a very special point of space, since the
density at this point blows up. It is called a `singularity' or a `hole'.
It is as though all the matter of the dead star fell through the
`hole'. Thus, a black hole has {\it no structure}! It is just pure
gravitation devouring anything around, with a special point
where all the stuff ends up, and a geometrical (not real) surface acting
like a one-way membrane hiding this special singular point from view of
external observers. 

There is however an error in all of this: from the Newtonian perspective,
photons are quite insensitive to gravitational forces. Our above
considerations had endowed them with a small mass, but that is not
quite right. To mediate long range electromagnetic forces as they do,
photons must have strictly zero mass. So what went wrong ? The point is
that we already know that this Newtonian perspective is itself limited; it
was replaced about 90 years ago by Einstein's General Relativity (GR).
According to GR, gravitation is no longer to be thought as a force, but
only as a manifestation of curvature of spacetime; free particles, both
massive and massless (like photons), in such a curved geometry are bound
to follow trajectories that are not straight as in flat (Minkowski)
spacetime, but bend around the source of the curvature. Also, given the
special relativistic equivalence of mass and energy, it stands to reason
that anything that has energy (and momentum, but no rest mass) can equally
generate a gravitational field as spacetime curvature. This idea is
enshrined in the celebrated Einstein equation of GR :
$Curvature ~\propto~energy-momentum~ density$ with Newton's gravitational
constant $G$ appearing in the proportionality constant.

How do black holes fit into GR ? They fit in very nicely, as {\it exact}
solutions to Einstein's beautiful but very complicated equation. They
describe spacetimes with curvature, but also possess the property
that
these spacetimes are {\it singular} - the curvature is infinite at a
particular point, consistent with the density blowing up as discussed
above. Secondly, the `trapped' surfaces beyond which nothing escapes
appear as null surfaces in these exact solutions - they are know as event
horizons. Finally, there is this amazing morphology of black holes in GR:
they are completely classified by three parameters - mass, electric charge
and angular momentum. The astrophysicist S. Chandrasekhar summarised these
properties rather eloquently when he described black holes as `the most
perfect macroscopic objects there are in the universe. ... the simplest as
well.'\cite{chandra}

\section{Laws of Black Hole Mechanics and Entropy}

This simple picture of black holes changed dramatically in the early and
mid-seventies. It all started with Hawking's discovery \cite{haw1} within
GR that the area of the event horizon of a black hole cannot decrease in
any physical process. In other words, if two black holes of horizon areas
${\cal A}_1$ and ${\cal A}_2$ coalesced into a single black hole, albeit
with release of a lot of gravitational radiation, the resultant black hole
cannot have an area less than ${\cal A}_1+{\cal A}_2$. This was followed
by the discovery \cite{bch}, also within classical GR, of two other laws,
the so-called Zeroth and First Laws of Black Hole Mechanics. The Zeroth
law states that the geometrical quantity known as surface gravity
(something like the Newtonian `acceleration due to the gravity') remains
constant on the horizon. The First law, for the Schwarzschild black hole
(which is completely characterised by mass $M$ with charge $Q=0$ and
angular momentum $J=0$), states that
\begin{equation} 
d M~=~{\kappa \over 2\pi}~d {\cal A}~,\label{firl} 
\end{equation}
where the surface gravity $\kappa \equiv 1/4M$ on the horizon. For black
holes with all three parameters non-zero (the Kerr-Newman family), the
first law takes the form
\begin{equation}
dM~\equiv~{\kappa \over 2\pi}~d {\cal A}~+~\Phi~dQ~+~{\vec \Omega}
\cdot
d{\vec L}~, ~\label{kern}
\end{equation}
where, $\kappa \equiv (r_+ - r_-)/4 {\cal A}~,~\Phi \equiv 4\pi Q
r_+ /{\cal
A}~,~{\vec \Omega}
\equiv 4\pi {\vec L}/M {\cal A}_{hor}~$. Hawking's earlier discovery of
horizon area non-decrease is called the Second Law of Black Hole
Mechanics. 

The analogy of these three laws with the Zeroth, First and Second laws of
ordinary thermodynamics is hard to miss, with the surface gravity being
analogous to temperature, horizon area with entropy and mass with
internal energy. Coupled with the usual description of the electrostatic
potential $\Phi$ and the angular velocity $\Omega$ on the horizon, the
second and third terms in (\ref{kern}) are together like the usual `$PdV$'
term of the First law of ordinary thermodynamics. But is this all there is
to these laws of black hole mechanics ? 

Bekenstein \cite{bek} was the first to go beyond mere analogizing and
propose that black holes actually {\it possess entropy} - a truly bold
hypothesis. He asserted that the horizon area was a measure of how much
entropy a black hole could have, in sharp contrast to standard
thermodynamic notions where entropy is supposed to be a function of
volume. Drawing upon work of Christodolou and others, Bekenstein concluded
that black hole entropy must be {\it proportional} to the horizon
area: $S_{bh} \propto {\cal A}$. In order that $S_{bh}$ be dimensionless,
the proportionality constant must have dimensions of inverse
squared
length. However, if this relation is to be universally valid for all black
holes, this constant should be independent of black hole parameters. Also,
it should not depend on interaction constants of non-gravitational
interactions. Thus, the only available fundamental length in this case is
the Planck length, $\ell_P \equiv (G \hbar /c^3)^{1/2} \sim
10^{-33}~cm$. Thus, 
\begin{equation}
S_{bh}~=~\eta ~{{\cal A} \over \ell_P^2}~, \label{bekl}
\end{equation}
where $\eta$ is a dimensionless constant of $O(1)$. Now, the Planck length
is the distance from a point particle, for instance, at which its
gravitational effect characterised by its Schwarzschild radius, becomes as
significant as its quantum effect characterised by its Compton
wavelength:
\begin{equation}
\ell_P ~\sim ~r_S ~\sim \lambda_C~, \label{pla}
\end{equation}
where, $r_S \equiv 2GM/c^2~,~\lambda_C \equiv \hbar/Mc$; elimination of the
mass $M$ from the second (approximate) inequality in (\ref{pla}) and
substituting back in either $r_S$ or $\lambda_C$ yields the standard
expression for $\ell_P$ given above. In this sense, this length typifies the
length scale at which
{\it quantum} gravitational effects can no longer be ignored. It follows
that, black hole entropy must therefore have origins that are actually
quantum gravitational in nature. Indeed, an object of pure gravity with no
structure classically should not possess any entropy at all. The only way
it could possess entropy is through microstates that appear in a truly
quantum description. 

Bekenstein's proposal of black hole entropy thus gives us a deep reason to
worry about the programme of quantization of gravitation - a programme
that has historically been steeped with difficulties and abandoned as
hopeless by most practitioners of GR, until relatively recently. But it
also raises fundamental issues:
\begin{itemize}
\item Why should black holes at all have entropy ?
\item In what sense is $T_{bh} \equiv \kappa/2\pi$ the temperature of the
black hole ?
\item What becomes of the standard Second law of thermodynamics in
presence of black holes ?
\item What are the microstates whose counting would yield the area law for
black hole entropy ?
\end{itemize}

We analyze these issues one by one. A gas in a container has a very large
number of molecules in random motion. A microscopic description of this
system in terms of canonical coordinates and momenta of each particle is
actually impossible. Thus, a statistical description, in terms of
average macroscopic variables like temperature, pressure, etc. is the best
one can have of such a system. This means that one has less than complete
information about the system, and this lack of information manifests as
entropy. In case of black holes, this lack of information is contained in
our lack of information about the very nature of gravitational
collapse. Thus, the parameters characterising black holes in GR actually
do not specify individual black holes; rather they, like the
temperature and pressure of a gas, are average parameters describing
equivalence classes of black holes, each of which may have collapsed from
a very different star via a very different process.

The temperature $T_{bh}$ is not the temperature of the horizon of the
black hole in the sense that a freely falling observer will sense it as
she crosses the horizon. At the horizon, the redshift factor vanishes, so
the observer detects no temperature at all. The notion of black hole
temperature is made unambiguous if we imagine placing the black hole in a
background of black body radiation in equilibrium at a temperature $T <
T_{bh}$. In his celebrated work, Hawking \cite{haw2} showed that, for an
observer located at infinity with a vanishing ambient temperature, a
black hole actually radiates in a
Planckian spectrum like a black body at a temperature $T_{bh}$ ! The
mean number of particles emitted at a frequency $\omega$ is given by the
Planckian formula
\begin{equation}
n_{\omega}~=~{|t_{\omega}|^2 \over {\exp (\hbar \omega/ T_{bh}) -1}}
~, ~\label{hrad}
\end{equation}
where, $t_{\omega}$ is the absorption coefficient and we have set the
Boltzmann constant $k_B=1$. The larger the mass of the black hole, the
smaller is this
equilibrium (or Hawking) temperature $T_{bh}$, so that for most stellar
black holes, this temperature $T_{bh} << 2.7 deg. K$, so that Hawking
radiation from such black holes is swamped by the cosmic microwave
background. In fact, these black holes absorb rather than emit,
radiation. One other outcome of Hawking's work is fixing the constant
$\eta=\frac14$ in (\ref{bekl}); this law, $S_{BH}={\cal A}/4 \ell_P^2$ is
called the Bekenstein-Hawking Area Law (BHAL). 

As for the status of the second law of thermodynamics in presence of
black holes, it
is true that when matter falls across the event horizon into a black
hole, the entropy of the matter is lost. One might think of this
as a decrease in the entropy of the universe. However, as Bekenstein
\cite{bek} pointed out, when matter falls into a black hole, the mass of
the black hole, and hence its area, increases. The hypothesis of black
hole entropy then says that the entropy of the black hole must increase as
a consequence, by at least the same amount as the entropy lost by the the
part of the universe outside the event horizon. Thus, in presence of black
holes, the Second law of thermodynamics is modified into the statement
that {\it the total entropy, defined as the sum of the entropy outside the
event horizon and the black hole entropy, can never decrease in any
physical process},
\begin{equation}
\delta (S_{ext}~+~S_{bh}) \geq 0 ~. \label{g2l}
\end{equation}
This relation is called the Generalized Second Law of thermodynamics and 
its enunciation marks the beginning of the subject known as Black Hole
Thermodynamics. 

\section{Microstates}

\subsection{Generalities}

We now come to the issue of microstates. As mentioned earlier, black hole
entropy originates from microstates that must be essentially quantum
gravitational in nature. This means that one must consider at least the
{\it kinematical} aspects of a quantum theory of gravitation. Since
entropy entails counting the degrees of freedom of the theory rather than
studying the dynamics of these degrees of freedom, a priori one may hope
that a kinematical approach may work, especially since there is no
complete theory of quantum gravitation yet at our disposal.  

Before subscribing to a particular proposal for a quantum GR, one may make
some progress in terms of very general pictures of what a black hole may
look like in quantum gravity. A crucial role is played by the Planck
length $\ell_P$, which is like $\hbar$ in the quantum theory of spacetime.
Thus, the spacetime continuum may degenerate, at such a length scale, to
some sort of a lattice structure with adjacent lattice sites separated by
this length scale. However, recall that classical GR has a huge symmetry -
those under coordinate diffeomorphisms in spacetime. Quantum GR should 
least retain at least some aspects of this symmetry, and from purely
kinematical
considerations, this might be taken to imply that there is no restriction
on the kind of lattice structure one should confine to. 

Furthermore, we are interested in a canonical picture, i.e., with what
happens at a {\it given} instant of time, and not the whole of spacetime.
To this effect, it is enough to deal with a {\it foliation} of a spacetime
containing a black hole, by a three dimensional spacelike hypersurface.
Such a foliation of the event horizon, which recall is a null hypersurface
which behaves as a one-way membrane, can be taken to be some (preferred)
spacelike two dimensional spheres. For what follows, this 2-sphere will
play the role of the horizon on a spatial slice. 

Now consider the same object within the very loose lattice structure
considered above. We can think of a two dimensional `floating' (as opposed
to a rigid) lattice basically covering the sphere. Since the area of a
tiny `plaquette' of this lattice may be taken to be Planck area
$\ell_P^2$, it stands to reason that the ratio of the macroscopic area of
the event horizon to the area of an elementary plaquette ${\cal
A}/\ell_P^2 >>1$. This latter inequality defines our notion of a
macroscopic black hole; our treatment in this review will focus on such
black holes, and not to those very interesting but difficult cases where
this number is $O(1)$. 

\subsection{It from Bit} 

Let us now place, following \cite{jaw}, binary variables (`bits') on the
lattice sites (or
equivalently, at the centre of the plaquettes). Then the number of such
variables $p \equiv \xi~{\cal A}/\ell_P^2~\gg~1$, where
$\xi=O(1)$. Without any loss of generality, $p$ can be taken to be an even
integer. Now, we can think of the two values of each binary variable as
characterising two quantum states, so that the size of the `Hilbert space'
of states on the (latticized) horizon is ${\cal N}_{bh} = 2^p$. The
entropy corresponding to this system is defined as $S_{bh} \equiv \log
{\cal N}_{bh}$; choosing the constant $\xi = (4 \log 2)^{-1}$, we obtain
\begin{equation}
S_{bh}~=~{{\cal A} \over \ell_P^2}~=~S_{BH}~.\label{bhal}
\end{equation}

The generality of the above scenario makes it appealing vis-a-vis a
quantum theory of black holes in particular and of quantum gravity in
general.  There is however one crucial aspect of any quantum approach to
black hole physics which seems to have been missed in the above, -- the
aspect of symmetry.  Indeed, the mere random distribution of spin 1/2
(binary) variables on the lattice which approximates the black hole
horizon, without regard to possible symmetries, possibly leads to a far
bigger space of states than the {\it physical} Hilbert space, and hence to
an overcounting of the number of the degrees of freedom, i.e., a larger
entropy.

\subsection {The physical Hilbert (sub)space} 

But what is the most plausible symmetry that one can impose on states so
as to identify the physical subspace ? The elementary variables are binary
or spin 1/2 variables which can be considered to be standard spin 1/2
variables under spatial rotations (more precisely $SU(2)$ doublets).
Recall now that at every point on a curved spacetime one can erect a local
Lorentz frame where the basic variables can be subjected to a local
Lorentz transformation. Here, of course, we are interested in a spatial
slice of the curved spacetime, and so the transformation of interest is a
`local spatial rotation'. Thus, the symmetry that one would like to impose
on the degrees of freedom obtained so far would be invariance under these
local spatial rotations in three dimensions. However, since one is dealing
with black holes of very large area, this amounts to considering `global'
or `rigid' spatial rotations or $SU(2)$ transformations. On very general
grounds then, the most natural symmetry of the physical subspace must be
this $SU(2)$ \cite{pm}. 

One is thus led to a symmetry criterion which defines the physical Hilbert
space ${\cal H}_S$ of horizon states contributing to black hole entropy:
{\it ${\cal H}_S$ consists of states that are compositions of
elementary $SU(2)$ doublet states with vanishing total spin ($SU(2)$
singlets)}. Observe that this criterion has no
allusions whatsoever to any specific proposal for a quantum theory of
gravitation. Nor does it involve any gauge redundancies (or any other
infinite dimensional symmetry like conformal invariance) at this point. It
is the most natural choice for the symmetry of physical horizon states
simply because in the `It from bit' picture, the basic variables are spin
1/2 variables. Later on we shall show
however that this symmetry arises very naturally in the Non-perturbative
Quantum GR approach known also as Quantum Geometry. It will emerge from
that approach that horizon states of large macroscopic black holes are
best described in terms of spin 1/2 variables at the punctures of a
punctured two-sphere which represents (a spatial slice of) the event
horizon.

The criterion of $SU(2)$ invariance leads to a simple way of counting the
dimensionality of the physical Hilbert space \cite{pm}. For $p$ variables,
this number is given by
\begin{equation}
dim{\cal H}_S \equiv {\cal N}(p)~=~\left( \begin{array}{c}
                         p \\ p/2
                        \end{array} \right)
                 ~ - ~\left( \begin{array}{c}
                         p \\ (p/2-1)
                         \end{array} \right)  ~\label{enpo}
\end{equation}
There is a simple intuitive way to understand the result embodied in
(\ref{enpo}). This formula counts the number of ways of making 
$SU(2)$ singlets from $p$ spin $1/2$ representations. The first term
corresponds to the number of states with net $J_3$ quantum number $m=0$
constructed by placing $m=\pm 1/2$ on the punctures.  However, this
term by itself {\it overcounts} the number of $SU(2)$ singlet states,
because even non-singlet states (with net integral spin, for $p$ is
an even integer) have a net $m=0$ sector. Beside having a sector with
total $m=0$, states with net integer spin have, of course, a sector with
overall $m=\pm 1$ as well. The second term basically eliminates these
non-singlet states with $m=0$, by counting the number of states
with net $m=\pm 1$ constructed from $m=\pm 1/2$ on the $p$
sites. The difference then is the net number of $SU(2)$ singlet
states that represents the dimensionality of ${\cal H}_S$. 

It may be pointed out that the first term in (\ref{enpo}) also has another
interpretation. It counts the number of ways binary variables
corresponding   
to spin-up and spin-down can be placed on the sites to yield a vanishing
total spin. Alternatively, one can think of the binary variables as unit
positive and negative $U(1)$ charges; the first term in (\ref{enpo}) then
corresponds to the dimensionality of the Hilbert space of $U(1)~invariant$
states. As already shown in \cite{dkm}, this corresponds to a {\it
binomial} rather than a random distribution of binary variables.

In the limit of very large $p$, one can evaluate the factorials in
(\ref{enpo})
using the Stirling approximation. One obtains
\begin{equation}
{\cal N}(p)~\approx~ {2^p \over p^{\frac32} }~. \label{plar}
\end{equation}
Clearly, the dimensionality of the physical Hilbert space is smaller than  
what one had earlier, as would be an obvious consequence of imposing
$SU(2)$ symmetry. Using the relation between $p$ and the classical horizon
area $A_S$ discussed in the last section, with the constant $\xi$ chosen
to 
take the same value as in that section, (\ref{plar}) can be shown
\cite{dkm} to lead to the following formula for black hole entropy,
\begin{equation}
S_{bh}~\equiv~\log {\cal N}(p)~\approx~  {{\cal A} \over{4\ell_P^2}}
~-~{3\over
2}~\log \left({{\cal A} \over{4\ell_P^2}} \right)~+~
const.~+~O({\cal A}^{-1}).
\label{main}
\end{equation}

The logarithmic correction to the BHAL is not unexpected if we think of
$S_{bh}({\cal A})$ as a power series for large ${\cal A}$ with $\frac14 {\cal
A}$ as the leading term; indeed, various approaches to computation of black
hole entropy (like the Euclidean path integral \cite{hawpa},  
Non-perturbative Quantum GR \cite{aa}, \cite{km}, boundary
conformal field theory \cite{car}, and so on) have been used, and a
general result like 
\begin{equation}
S_{bh}({\cal A})~=~~({{\cal A} \over 4\ell_P^2})~+~C~\log ( {{\cal A}
\over 4\ell_P^2})~, \label{genl}
\end{equation}
has been found, with various values of $C$, both positive and
negative. In some of the perturbative approaches, there is an added
constant $C' \sim
\log(\Lambda)$ where $\Lambda$ is a length cut-off needed to yield a finite
result for $S_{bh}$ \cite{mann}. This is quite in contrast to our result
(\ref{main}) where $S_{bh}$ is {\it intrinsically finite}. Note also that
according to the Second Law of Black Hole Mechanics, if two black holes
coalesce, the minimum area of the resultant black hole is the sum of the two
horizon areas. For such a coalescence, it is easy to see that, for $C>0$, 
\begin{equation}
S_{bh}({\cal A}_1~+~{\cal A}_2)~<~S_{bh}({\cal A}_1)~+~S_{bh}({\cal A}_2)~.
\label{cgt0}
\end{equation}
Assuming isolated eternal black holes which coalesce adiabatically with no
emission of gravitational waves, this property is perhaps not too
desirable from the point of view of the Second Law of Thermodynamics. From
this point of view also our result $C=-\frac32$ appears more
preferable. This is precisely the result that was obtained earlier from
Non-perturbative Quantum GR (also called Quantum Geometry) \cite{km} on
the basis of incipient contributions in ref. \cite{aa}. 

Even when the resultant black hole has a horizon area larger than the sum of
the areas, this preference for $C<0$ seems to hold, although in a slightly
weaker form. Also, from the theory of isolated horizons \label{aa1} which
incorporates radiation present in the vicinity of the horizon without crossing
it, this result seems to have a greater appeal than those others with $C>0$.

\section{Holography and the entropy bound}

Having identified the kinematical quantum states characterising a black hole
horizon, the question that immediately comes to mind is whether there are
other states that describe black hole physics. Although the `It from bit'
picture tends to imply that the entire information lies with the horizon
states, this has been more sharply articulated in the so-called Holographic
hypothesis \cite{thf}. According to this hypothesis, the horizon states
exhaust the Hilbert space of a black hole, encoding the entire information of
gravitationally collapsed matter in terms of macroscopic observables like
the horizon area. 

While a rigourous proof of this hypothesis is not available yet for
asymptotically flat spacetimes\footnote{For anti-de Sitter spacetimes the
so-called adS-CFT Correspondence conjecture \cite{mal} makes this
hypothesis plausible, although the conjecture itself warrants a proof.},
the beginnings of a general proof should first demonstrate the primacy of
quantum gravitational states associated with some spacetime boundary
rather than the quantum states describing bulk spacetime. Let us see
heuristically how this may come about.

Assume that the Hilbert space ${\cal H}$ of quantum GR in the presence of an
inner boundary (and the usual outer boundaries at asymptotic null infinity)
can be decomposed into states describing the geometry of `bulk' spacetime and
the boundary,
\begin{equation}
{\cal H} ~\sim~ {\cal H}_{bulk} \otimes {\cal H}_{bdy}~.
\end{equation}
The canonical partition function $Z(\beta)$ can be written formally as
\begin{equation}
Z(\beta)~=~Tr~\exp \{ -\beta~{\hat {\bf H}} \}~, \label{cpf}
\end{equation}
where, ${\hat {\bf H}}$ is the full Hamiltonian operator of quantum
gravitation. Now, we know that, in the canonical description of classical GR
\cite{aa2}
\begin{equation}
{\bf H}~=~{\bf H}_{bulk}~+~{\bf H}_{bdy}~ , \label{split}
\end{equation}
where, 
\begin{equation}
{\bf H}_{bulk}~ \approx ~ 0~, \label{const}
\end{equation}
being a sum of first class constraints generating general coordinate
transformations and internal (local Lorentz) gauge transformations on the
basic canonical variables (metric or tetrad, connections) of GR. The boundary
Hamiltonian ${\bf H}_{bdy}$ would, in the case of asymptotically flat
spacetime, give the ADM mass etc., and so does not vanish. 

Let us assume that the decomposition of the Hamiltonian (\ref{split}) and the
constraint relation (\ref{const}) hold for the full quantum gravitational
operator Hamiltonian, i.e., 
\begin{equation}
{\hat {\bf H}}_{bulk}~ |\Psi_{bulk} \rangle ~=~0~, \label{consq}
\end{equation}
where, $|\Psi_{bulk} \rangle \in {\cal H}_{bulk}$. The partition function
(\ref{cpf}) can now be written
\begin{eqnarray}
Z(\beta)~&=&~Tr_{bulk}~\exp \{-\beta ~{\hat {\bf H}}_{bulk} \} ~\cdot
~Tr_{bdy}~\exp \{-\beta~ {\hat {\bf H}}_{bdy}~ \}. \nonumber \\
&\equiv&~Z_{bulk}(\beta)~\cdot Z_{bdy}(\beta)~.\label{zsp}
\end{eqnarray}
Observe that $Z_{bulk}$ can be rewritten
\begin{eqnarray}
Z_{bulk}(\beta)~&=&~\sum_{\Psi_{bulk}}~\langle \Psi_{bulk}|\exp
\{ -\beta~{\hat {\bf H}}_{bulk} \} | \Psi_{bulk} \rangle~ \nonumber \\
&=&~dim~{ {\cal H}}_{bulk}~, \label{blpf}
\end{eqnarray}
using the operator constraint eq. (\ref{consq}). While the dimensionality
of ${\cal H}_{bulk}$ can indeed be infinity, the important point is that
it is a constant {\it independent of the inverse temperature $\beta$}. One
thus has
\begin{equation}
Z(\beta)~=~dim~{\cal H}_{bulk}~\cdot~Z_{bdy}(\beta)~. \label{hol}
\end{equation}
It follows that the entire non-trivial dependence on $\beta$ (and
thermodynamic implications thereof) of the partition function lies in the
`boundary' part of the partition function;  the bulk part merely
contributes an overall constant scale factor (an additive constant in the
free energy) which is not of much relevance thermodynamically.

Notice that the behaviour depicted in eq. (\ref{hol}) extends to the
situation with matter as well, since even in that case, without making any
approximations, one expects the quantum constraint (Wheeler-de Witt
equation) (\ref{consq}) to hold. So, within the tenets of non-perturbative
quantum GR (or quantum geometry) in the presence of an inner boundary like
a black hole horizon, non-trivial information appears to remain confined
to the boundary rather than the bulk. The situation changes dramatically
when ${\hat {\bf H}}_{bulk}$ is approximated by assuming a fixed classical
background and the quantum degrees of freedom are assumed to be small
perturbative fluctuations about this background, like in a theory of
gravitons. The full Wheeler-de Witt equation (\ref{consq}) is then
replaced by an inhomogeneous equation which exhibits only the linearized
symmetries at the quantum level to leading order. Under such conditions
the primacy of the boundary states vis-a-vis physical information is no
longer obvious. This is the situation, for instance, in (perturbative)
string theory which is basically a theory of gravitons. The bulk
partition function in this case is quite non-trivial. The plausibility
argument given above must therefore be relegated to the status of an
unproven hypothesis, in order that it can have physical applications, like
the calculation of black hole entropy. In contrast, a relation like
(\ref{hol}) appears to demystify to a large extent why black hole entropy
must be a function of horizon area rather than some bulk property. 

Let us now return to the issue of black hole entropy. The above discussion
of holography seems to justify focusing on the surface degrees of freedom
alone in the It from Bit picture and the subsequent Physical State
Criterion. Thus, black hole entropy can be taken to represent the {\it
maximal} possible entropy of a spacetime whose spatial slice has a
boundary that coincides with the intersection of this spatial slice with
the horizon. Now, it can be shown \cite{dkm} that eq.  (\ref{main})
actually translates into a bound on black hole entropy, given
by
\begin{equation}
S_{max}~=~\log \left( {\exp{S_{BH}} \over S_{BH}^{3/2} } \right)~.
\label{newb1}
\end{equation}

Thus, it follows that {\it all} 3-spaces with boundary have an entropy
bounded from above by (\ref{newb1}). That this is extremely plausible
follows from the following argument, based on {\it reductio ad absurdum}
\cite{smo}:  we assume, for simplicity that the spatial slice of the
boundary of an asymptotically flat spacetime has the topology of a
2-sphere on which is induced a spherically symmetric 2-metric. Let this
spacetime contain an object whose entropy exceeds the entropy bound given
in eq. (\ref{newb1}). Certainly, such a spacetime cannot have a black hole
horizon as a boundary, since then, its entropy would have been subject to
(\ref{newb1}). But, in that case, its energy should be less than that of a
black hole which has the 2-sphere as its horizon.  Let us now add energy
to the system, so that it does transform adiabatically into a black hole
with the said horizon, but without affecting the entropy of the exterior.
But we have already seen above that a black hole with such a horizon must
obey the bound (\ref{newb1}); it follows that the starting assumption of
the system having an entropy exceeding the bound (\ref{newb1}) must be
incorrect. 

Thus, we have indeed obtained an upper bound on the entropy of a large
class of spacetimes.  Notice that this bound {\it tightens} the
semiclassical Bekenstein bound \cite{bek}, which is of course expected
because of its quantum kinematical underpinning. We now turn to a brief,
mostly qualitative, expos\'e of aspects of Non-perturbative Quantum GR
(also called Quantum Geometry) which is a candidate theory that provides
such underpinning. 

\section{Entropy from Quantum Geometry}

Quantum Geometry (QGeo) is a framework for non-perturbative canonical
quantization of four dimensional general relativity, consistent with all
symmetries of the latter. These symmetries include general coordinate
transformations and local Lorentz transformations; within GR these
symmetries are generated by first class constraints just as in Maxwell
electrodynamics the Gauss law constraint generates (time-independent) 
gauge transformations which are symmetries of the theory. In the quantum
theory, one expects that the Hilbert space will be spanned by states that
are annihilated by the constraints, now appearing as quantum mechanical
operators. Thus, one could attempt to find the Hilbert space by solving
these operator constraints. In practice, this turns out to be a formidable
task and all parts of this programme have not been completed yet for QGeo.
One finds a {\it kinematical} Hilbert space whose elements solve some of
the constraints, but not all. As we have already mentioned, this
kinematical Hilbert space turns out to be sufficient for the calculation
of black hole entropy.

What sort of wave-functions span the kinematical Hilbert space ? We know
for example that the configuration space of classical scalar field theory
in Minkowski spacetime consists of square integrable function(al)s of
smooth scalar fields. However, the situation changes dramatically for the
quantum configuration space where {\it non-smooth} Dirac delta
function-lke distributions must also be included in addition to smooth
scalar fields. Similarly, the configuration space of classical GR consists
of smooth metrics, but the wave functions of QGeo are function(al)s of
metrics that are not necessarily smooth. Thus, one has to consider metrics
(and hence curvatures) which are actually generalized functions of
spacetime coordinates (i.e., like Dirac delta functions), being
non-vanishing only over certain subspaces of the classical continuum.

It turns out that a good basis for the kinematical Hilbert space of QGeo
is the so-called Spin Network basis. Typically, a spin network is
a three
dimensional `floating' lattice embedded in the three dimensional spacelike
hypersurface that slices spacetime. It consists of links each carrying a
{\it spin} $j$ which can take integral as well as half-odd integral
values. These links meet at vertices where one inserts certain tensors
which are invariant under local spatial rotations.  Specifying the spins
on all links and the invariant tensors at all vertices gives a complete
specification of the spin network state. No links are left sticking out of
the network, so that the net spin of the state vanishes. Also, the lengths
of links are quite immaterial for all physical considerations. These
properties ensure that the state is invariant under local spatial
rotations and time independent coordinate transformations respectively.
However, whether it is invariant under time dependent coordinate
transformations is not yet clear. States constructed out of such networks
form a complete orthonormal basis for the kinematical Hilbert space. Now,
from a classical geometrical standpoint, the curvature is non-vanishing
only on the network - it is non-smooth in that sense. 

What are the observables in this framework ? These are geometrical in
nature, like length of a curve, area of a two-surface, volume of a
three-surface etc. The spin network basis turns out to be an eigenbasis
for these obervables. For instance, a two-surface $S$ with classical area
$A$ (w.r.t. some classical metric) has a well-defined area operator ${\hat
A}$ associated with it, such that the spin network basis is an eigenbasis
of this operator, with a {\it discrete} spectrum \cite{al}. One can think
of placing this two-surface inside the spin network such that it is
pierced by the links of the network. Thus the two-surface consists of a
two dimensional lattice of punctures, with spins $j_i$ at the $i$th
puncture, $i=1,2, \dots, p$. Again, these punctures correspond to deficit
angles in classical geometrical terms, emphasizing the non-smoothness of
the induced metric on the two-surface. In terms of these spins, the area
eigenvalues lying within a Planck area of the classical area are given
by 
\begin{equation}
{\cal A} ~=~8\pi ~\gamma \ell_P^2~\sum_{i=1}^p~\sqrt{j_i(j_i+1)}~,
\label{area}
\end{equation}
where, $\gamma$ is the so-called Barbero-Immirzi parameter characterising
inequivalent quantizations of the same classical theory, something akin
to the $\theta$ parameter in Yang-Mills theory. 

In this framework, it is not yet possible to describe black holes as
quantum states that solve all operator constraints. Instead, the idea is
to treat the classical event horizon as an inner boundary of spacetime
which is canonically quantized using the spin network basis. The boundary
conditions imposed on this inner boundary of course correspond strictly to
those on the black hole horizon. Very detailed analyses of these boundary
conditions have been performed \cite{aa3}. These boundary conditions
require that the gravitational action be augmented by the action of a
Chern-Simons theory living on the horizon.  Boundary states of the
Chern-Simons theory constitute precisely the microstates that contribute
to the entropy. These states correspond to conformal blocks of the
two-dimensional Wess-Zumino model that lives on the spatial slice of the
horizon, which is a 2-sphere of area ${\cal A}$.  The dimensionality of
the boundary Hilbert space has been calculated thus \cite{km}
by counting the number of conformal blocks of two-dimensional $SU(2)_k$
Wess-Zumino model, for arbitrary level $k$ and number of punctures $p$ on
the 2-sphere. The Chern Simons coupling constant $k \sim {\cal A}$. We
shall show, from the formula for the number of conformal blocks
specialized to macroscopic black holes characterized by large $k$ and $p$
\cite{km}, that eq.  (\ref{main}) ensues.

Let us start with the formula for the number of conformal blocks of
two-dimensional $SU(2)_k$ Wess-Zumino model that lives on the punctured
2-sphere. For a set of punctures ${\cal P}$ with spins $ \{j_1, j_2, \dots
j_p \} $ at punctures $\{ 1,2, \dots, p \}$, this number is given by
\cite{km}, for large $k \rightarrow \infty$
\begin{equation}
N^{\cal P}~=~\sum_{m_1= -j_1}^{j_1} \cdots \sum_{m_p=-j_p}^{j_p}  
\left[
~{\delta}_{(\sum_{n=1}^p m_n), 0}~-~\frac12~
{\delta}_{(\sum_{n=1}^p m_n),
1}~-~
\frac12 ~{\delta}_{(\sum_{n=1}^p m_n), -1} ~\right ]. \label{kinf}
\end{equation}

We consider a large fixed classical area of the horizon,
and ask what the largest value of number of punctures $p$ should be,
so as to be consistent with (\ref{area}); this is clearly obtained when
the spin at {\it each} puncture assumes its lowest nontrivial value of
1/2, so that, the relevant number of punctures $p$ is given by
\begin{equation}
p~=~{{\cal A} \over 4 \ell_{P}^2}~{\gamma_0 \over \gamma}~, \label{pmax}
\end{equation}
where, $\gamma_0=1/\pi \sqrt{3}$. We are of course interested in the case
of very large $p$ appropriate to a macroscopic black hole. 

Now, with the spins at all punctures set to 1/2, the number of states for
this set of punctures ${\cal P}$ is given by
\begin{equation}
N^{{\cal P}}~=~\sum_{m_1= -1/2}^{1/2} \cdots \sum_{m_{p}=-1/2}^{1/2}
\left[ ~\delta_{(\sum_{n=1}^{p} m_n), 0}~-~\frac12~
\delta_{(\sum_{n=1}^{p} m_n),
1}~-~\frac12 ~\delta_{(\sum_{n=1}^{p} m_n), -1} ~\right ] \label{excto}
\end{equation}
The summations can now be easily performed, with the result given 
precisely by the {\it rhs} of eq. (\ref{main}).

This establishes on a microscopic basis the validity of the extension of
the `It from bit' picture proposed by us earlier. The central
role played by variables in the doublet representation of a (global)
$SU(2)$ group, which we identified with the binary variables on the
lattice approximating the horizon, is now clarifed. This completes the
derivation of our physical space criterion and the ensuing entropy formula
and holographic bound on the basis of a quantum kinematical formulation. 

\section{Conclusion}

Quantum Geometry as a non-perturbative approach to quantum GR has had a
measure of success in providing an ab initio way to calculate the entropy
of generic non-rotating four dimensional black holes. The results obtained
within this formalism establish on firm foundations heuristic results
that can be obtained using somewhat more intuitive arguments. That all the
seemingly disparate elements of the formalism converge to these results
points to their inherent robustness. 

It would be most gratifying to have a complete rigourous proof of the
so-called holographic hypothesis, so germane to all considerations of
entropy of spacetimes with horizons. Hopefully, we have been able to
provide a reasonable intuitive argument as to how such a proof might go. A
test of the approach given here would be the derivation of logarithmic
corrections to black hole entropy that we have found, from a path integral
framework, using only the boundary action and fluctuations around its
classical value (which is known to yield the BHAL in the Euclidean
approach).

\end{document}